\documentclass[10pt,twocolumn,letterpaper]{article}

\usepackage[pagenumbers]{wacv}

\definecolor{wacvblue}{rgb}{0.21,0.49,0.74}
\usepackage[pagebackref,breaklinks,colorlinks,allcolors=wacvblue]{hyperref}

\usepackage{makecell}

\newcommand{\model}{\epsilon_\theta}

\newcommand{\conditioner}{\tau_\xi}

\newcommand{\clipimgenc}{\tau_\phi}

\newcommand{\expec}{\mathbb{E}}

\newcommand{\textprompt}{\mathcal{P}}
\newcommand{\textembedding}{\mathcal{C}_{text}}
\newcommand{\ipimgcond}{\mathcal{C}_{img}}
\newcommand{\ipcondmlp}{\mathbf{IP}}

\newcommand{\sdxl}{\mathcal{G}}

\newcommand{\ourmethod}{\textit{ColorWave }}

\newcommand{\ipcondimg}{\mathcal{I}_{ip}}

\newcommand{\minisection}[1]{\vspace{0.02in}\noindent{\bf #1}}

\newcommand{\quotes}[1]{``#1''}
\newcommand\blfootnote[1]{
  \begingroup
  \renewcommand\thefootnote{}\footnote{#1}
  \addtocounter{footnote}{-1}
  \endgroup
}
\usepackage{multirow}

\def\wacvPaperID{1301}
\def\confName{WACV}
\def\confYear{2026}

\title{Leveraging Semantic Attribute Binding for \\ Free-Lunch Color Control in Diffusion Models}

\author{
Héctor Laria$^{1,2}$,
Alexandra Gomez-Villa$^{1,2}$,
Jiang Qin$^{3}$,
Muhammad Atif Butt$^{1,2}$,\\
Bogdan Raducanu$^{1,2}$,
Javier Vazquez-Corral$^{1,2}$,
Joost van de Weijer$^{1,2}$,
Kai Wang$^{1,4,5,\thanks{Corresponding author.}}$\\
\\
$^{1}$Computer Vision Center, Spain
$^{2}$Universitat Autònoma de Barcelona\\
$^{3}$Harbin Institute of Technology
$^{4}$City University of Hong Kong \\
$^{5}$Program of Computer Science, City University of Hong Kong (Dongguan)
}

\begin{document}

\twocolumn[{
\renewcommand\twocolumn[1][]{#1}
\maketitle
\begin{center}
    \centering
    \captionsetup{type=figure}
    \includegraphics[width=\textwidth]{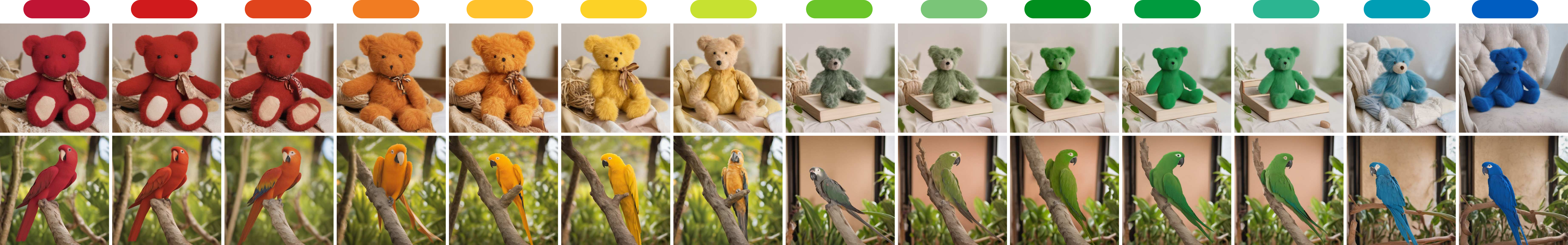}
    \vspace{-6mm}
    \captionof{figure}
    {
    \ourmethod accurately reproduces subtle color variations in smooth interpolation between similar tones. Each column shows a different target object rendered with gradually shifting colors (displayed above each image). The results demonstrate that our method \ourmethod is sensitive to small changes in the RGB color space while preserving realistic object appearance and scene composition.
    }
    \label{fig:color_interpolation}
\end{center}
}]

\begin{abstract}
Recent advances in text-to-image (T2I) diffusion models have enabled remarkable control over various attributes, yet precise color specification remains a fundamental challenge. Existing approaches, such as ColorPeel, rely on model personalization, requiring additional optimization and limiting flexibility in specifying arbitrary colors. In this work, we introduce ColorWave, a novel training-free approach that achieves exact RGB-level color control in diffusion models without fine-tuning. By systematically analyzing the cross-attention mechanisms within IP-Adapter, we uncover an implicit binding between textual color descriptors and reference image features. Leveraging this insight, our method rewires these bindings to enforce precise color attribution while preserving the generative capabilities of pretrained models. Our approach maintains generation quality and diversity, outperforming prior methods in accuracy and applicability across diverse object categories. Through extensive evaluations, we demonstrate that \ourmethod establishes a new paradigm for structured, color-consistent diffusion-based image synthesis.
\vspace{-2mm}
\blfootnote{\hspace{-5.8mm} Corresponding Author (*): Kai Wang (kai.wang@cityu-dg.edu.cn) \\
Project page: {\scriptsize \url{https://hecoding.github.io/colorwave-page}}}
\end{abstract}

\section{Introduction}

Text-to-image (T2I) diffusion models~\cite{deepfloyd,Rombach_2022_CVPR_stablediffusion,saharia2022imagen} have revolutionized image generation, offering unprecedented control over composition~\cite{liu2022compositionaldiffusion,wang2024compositional,wang2023dynamic}, style~\cite{sohn2023styledrop,wang2024instantstyle,xing2024csgo,qin2025freelunch}, subject matter~\cite{textual_inversion,ruiz2022dreambooth}, and visual aesthetics~\cite{gomez2024art_illusion,geng2024visual_anagrams}. Among these attributes, color is particularly crucial, fundamentally shaping viewer perception, conveying emotional tone, and serving as a critical design specification in professional contexts from brand identity to product development~\cite{singh2006impact}.
Despite remarkable progress, T2I models exhibit significant limitations in precise color control. When users specify colors through linguistic descriptors like \quotes{red} or \quotes{blue}, these terms encompass broad ranges of potential shades, making exact color matching challenging—a capability essential for design applications where color fidelity is non-negotiable.

T2I personalization methods such as DreamBooth~\cite{ruiz2022dreambooth} or CustomDiffusion~\cite{kumari2022customdiffusion} can address this limitation but often fail to correctly disentangle shape from color information~\cite{butt2025colorpeel}. ColorPeel~\cite{butt2025colorpeel} enables precise color control by learning specific color prompts through additional training. However, these approaches require separate optimization processes for each individual color, resulting in significant computational overhead and limiting practicality. The need for model fine-tuning creates barriers to flexibly specifying arbitrary colors during inference time. Furthermore, learning-based strategies become increasingly impractical in advanced T2I models with multiple text encoders such as SDXL~\cite{podell2023sdxl}, SD3~\cite{esser2024sd3}, and FLUX~\cite{flux2024}.

Recent advancements in image conditioning for diffusion models, particularly through mechanisms like IP-Adapter~\cite{ye2023ip,wang2024instantstyle,xing2024csgo}, have enabled new forms of control by conditioning generation on reference images. While these adapters were primarily designed for style transfer or subject-driven generation, they encode complex visual information in ways that have not been fully explored. Through systematic investigation, we discovered a previously unrecognized property: \emph{semantic attribute binding} which refers to the implicit bindings between visual attributes and their semantic representations within multi-modal space.
Specifically, we found that these bindings create a mapping between RGB values in reference images and linguistic color descriptors in text prompts—a connection that existing methods, including IP-Adapter itself, do not explicitly exploit. Our insight enables precise RGB-level color manipulation without any model fine-tuning, revealing capabilities latent within IP-Adapter that haven't been previously identified or utilized for targeted control.

In this paper, we introduce \textit{ColorWave}, a novel training-free approach for precise color control in diffusion models. Unlike previous methods, \ourmethod enables users to specify exact RGB values (see Figure \ref{fig:color_interpolation}) for objects in generated images without requiring any additional training or model fine-tuning. Our method leverages  semantic attribute binding capabilities within the IP-Adapter framework, effectively \quotes{rewiring} these connections to achieve precise color attribution to target objects. By introducing minimal additional modules that exploit pretrained diffusion models, we achieve unprecedented color accuracy while maintaining generation quality and diversity.
To summarize, our key contributions include:
\begin{itemize}
    \item The \textit{first} training-free approach \ourmethod for precise color control in diffusion-based generation, enabling specification of arbitrary colors without additional optimization.
    \item A novel technique that exploits the binding mechanism between visual features and semantic counterparts in the latent space of diffusion models.
    \item A selective attention modulation strategy that preserves the generative capabilities of the base model while enabling precise color control for targeted objects.
    \item Comprehensive evaluations demonstrating superior performance across diverse object categories and color specifications compared to existing methods, both in terms of color accuracy and generation quality.
\end{itemize}

\section{Related work}

\noindent \textbf{Image-Conditioned T2I Diffusion Models.}
Conditional image-based T2I models integrate auxiliary image inputs with textual prompts, allowing refined control over spatial layout~\cite{Li2024ControlNet++:Feedback,Choi2023FineControlNet:Injection}, object content~\cite{chen2023suti,li2024blipdiffusion}, and stylistic attributes~\cite{sohn2023styledrop,han2024stylebooth}.
ControlNet~\cite{zhang2023controlnet} employs a duplicated U-Net~\cite{ronneberger2015unet} encoder for pixel-level structural guidance, while UniControlNet~\cite{zhao2023unicontrolnet} extends this with dual adapters handling both local and global controls.

As another technique pipeline, IP-Adapter~\cite{ye2023ip} leverages the vision encoder from CLIP~\cite{radford2021clip} to extract image features, integrating them into the U-Net backbone via cross-attention layers for more coherent conditioning.
Building on this, IPAdapter-Instruct~\cite{rowles2024ipadapterinstruct} introduces an instructional prompt mechanism, enhancing interpretability and reducing the need for separate model training for different conditioning inputs, addressing a key limitation of ControlNet.
However, the role of the additional cross-attention mechanism in IP-Adapter remains underexplored. In this paper, we present the first findings on its properties, revealing that IP-Adapter implicitly establishes a binding between visual attributes and their corresponding semantic representations.

\noindent \textbf{Color Control in T2I diffusion models.}
With the rapid advancements in T2I generation,
various text-guided image editing approaches~\cite{hertz2023delta_DDS,meng2022sdedit,mokady2022null,zhang2023forgedit,chen2023fec,li2023stylediffusion,tang2023iterinv} have been developed to enable controllable modifications. For instance, methods like Imagic~\cite{kawar2022imagic} and P2P~\cite{hertz2022prompt} leverage Stable Diffusion (SD) models for structure-preserving edits.
Another technique stream which can also achieve controllable generation is transfer learning for T2I models~\cite{ruiz2022dreambooth,kumari2022customdiffusion,wang2024mcti,textual_inversion}, or also referred to
\textit{personalized generation}.
It aims at adapting a given model to a \textit{new concept} by given images from the users and bind the new concept with a unique token. As a result, the adaptation model can generate various renditions for the new concept guided by text prompts.

However, all these existing techniques rely heavily on the generative capacity of diffusion models and struggle to achieve fine-grained control over color attributes in image editing and generation tasks. Only a limited number of works~\cite{butt2025colorpeel,ge2023richtext} have begun addressing the challenge of precise color generation. Rich-Text~\cite{ge2023richtext} enhances color fidelity through multi-pass processing with global and local diffusion models, but at the cost of high computational overhead and reduced color accuracy. In contrast, ColorPeel~\cite{butt2025colorpeel} introduces color prompt learning to improve color alignment with user input. However, this method requires extensive training and is restricted to handling only a few color names per training session.
In this paper, we argue that the limitations of prior approaches stem from their failure to recognize the implicit attribute binding between the IP-Adapter~\cite{ye2023ip} module and T2I diffusion models~\cite{podell2023sdxl,Rombach_2022_CVPR_stablediffusion}. Leveraging this property, we propose a training-free solution for precise color generation, allowing seamless adaptation to any user-specified color inputs, particularly benefiting artistic and creative applications.
Our work reveals that precise color control capabilities already exist within standard IP-Adapter architectures through semantic attribute binding, eliminating the need for additional training while achieving competitive results.

\section{Methodology}
\subsection{Preliminaries}

\minisection{T2I Diffusion Models.}
We built on the SDXL~\citep{podell2023sdxl} model, consisting of two primary components: an autoencoder
and a diffusion model $\model$.
The model $\epsilon_{\theta}$ is trained by the loss:
\begin{equation}
L_{LDM} = \expec_{z_0, \epsilon \sim \mathcal{N}(0, 1), t }\Big[ \Vert \epsilon - \model(z_{t},t, \conditioner(\textprompt)) \Vert_{2}^{2}\Big]
\label{eq:ldm_loss}
\end{equation}
where  $\model$ is a  UNet, conditioning a latent input $z_{t}$, a timestep $t \sim \text{U}(1,T)$, and a text embedding $\conditioner(\textprompt)$.
More specifically, text-guided diffusion models generate an image from the textual condition as
$\textembedding=\conditioner(\textprompt)$,
where $\conditioner$ is the CLIP text encoder~\citep{Ilharco_Open_Clip_2021}
\footnote{SDXL uses two text encoders and concatenate the embeddings.}.
The cross-attention map is derived from $\model(z_t,t,\textembedding)$.
After predicting the noise, diffusion schedulers~\cite{Lu2022DPM-Solver++:Models,Lu2022DPM-Solver:Steps} are used to predict the latent $z_{t-1}$. As an example with the DDIM scheduler~\cite{Song2020DenoisingModels}, the formula is:
\begin{equation}\label{eq:ddim_sampling}
\resizebox{0.999\linewidth}{!}{$
\boldsymbol{z_{t-1}} = \sqrt{\frac{\alpha_{t-1}}{\alpha_t}}\boldsymbol{z_{t}} + \sqrt{\alpha_{t-1}}\left(\sqrt{\frac{1}{\alpha_{t-1}}-1}-\sqrt{\frac{1}{\alpha_t}-1}\right) \cdot \model(z_t,t,\textembedding)
$}
\end{equation}
where $\alpha_t$ is a predefined scalar function. Here we simplify the $z_{t-1}$ inference process as $z_{t-1}=\sdxl (z_t,t,\textembedding)$.

\minisection{IP-Adapter.}
Building on T2I diffusion models,
the IP-Adapter~\cite{ye2023ip} introduces additional controllability by conditioning the T2I model on a conditional image $\ipcondimg$. Practically, this involves leveraging a pretrained T2I diffusion model and incorporating a cross-attention layer to the (projected) image condition following each text-prompt conditioning layer.
The conditional image is encoded in the low-dimensional CLIP image embedding space~\cite{Ilharco_Open_Clip_2021} to capture high-level semantic information.
By denoting the CLIP image encoder as $\clipimgenc$ and IP-Adapter projection as $\ipcondmlp$, this process is adding a new image condition $\ipimgcond=\ipcondmlp(\clipimgenc(\ipcondimg))$ to the T2I model as $z_{t-1}=\sdxl (z_t,t,\textembedding,\ipimgcond)$.

\begin{figure}[t]
  \centering
   \includegraphics[width=\linewidth]{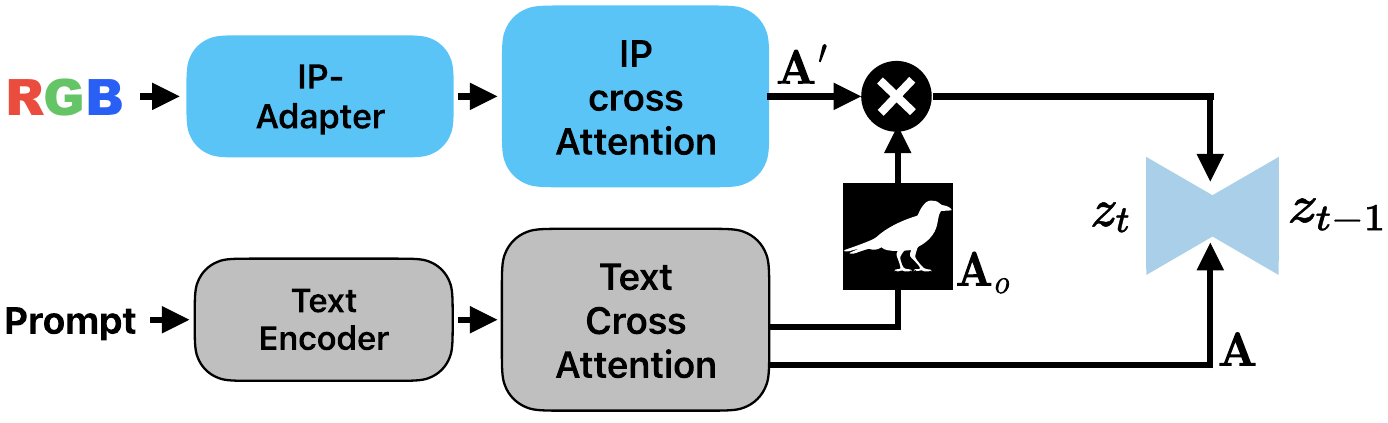}
   \vspace{-5mm}
   \caption{\textbf{Overview of \textit{ColorWave}}. Our approach leverages semantic attribute binding between IP-Adapter and text cross-attention pathways to achieve precise color control. User-specified RGB values are encoded through IP-Adapter and selectively bound to object tokens in the text prompt, enabling training-free color attribution while preserving generative quality.}
   \label{fig:overview}
   \vspace{-4mm}
\end{figure}

\minisection{Studied architecture.}
An overview of our approach, \textit{ColorWave}, is provided in Figure~\ref{fig:overview}. At its core, our method leverages the base architecture of a pretrained diffusion model (SDXL) with an integrated IP-Adapter. The primary insight driving our work is the discovery of \textit{semantic attribute binding}—a phenomenon where visual attributes in reference images form implicit connections with their corresponding linguistic descriptors in text prompts. We demonstrate how this previously unexplored property can be harnessed for training-free color control in generated images.

To fully exploit this binding mechanism and achieve precise color control, \ourmethod introduces two key enhancements to the standard IP-Adapter framework: (1) automatic color name generation, which determines the optimal linguistic color descriptor for any user-specified RGB value; and (2) spatial prior addition, which ensures accurate color attribution to target objects by refining attention maps.
These components work in concert, effectively \quotes{rewiring} the cross-attention mechanism to strengthen the binding between user-specified colors and target objects while maintaining the generative capabilities of the underlying diffusion model. We detail \ourmethod in the following sections.

\subsection{Semantic attribute binding in IP-Adapters}\label{sec:SAB}

\begin{figure*}[ht]
\centering
\includegraphics[width=0.95\linewidth]{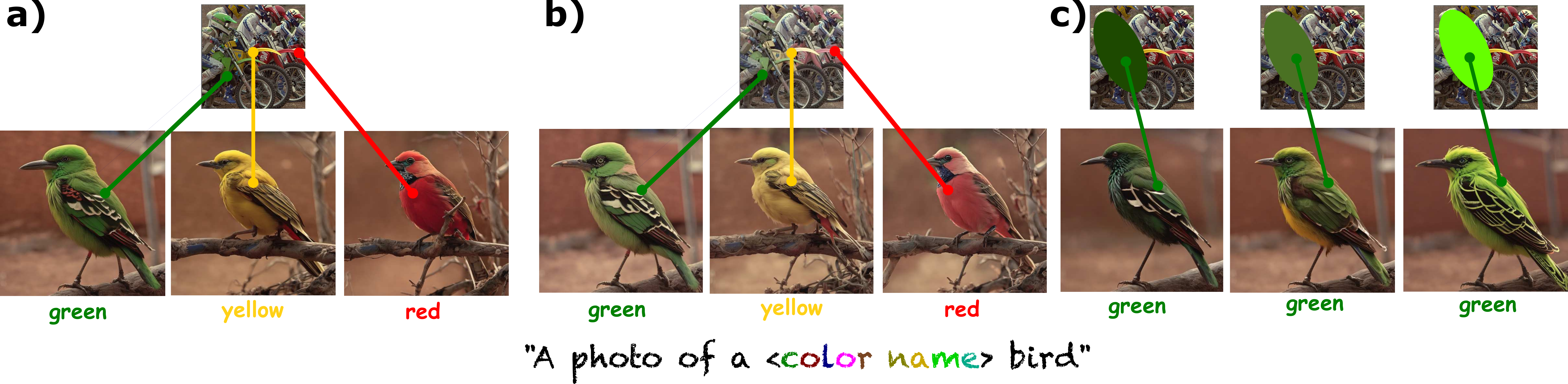}
\vspace{-3mm}
\caption{\textbf{Illustration of \emph{semantic attribute binding}}. On top, the color-guidance image is provided. a) Given the color name used in the prompt, the generated results will pick the respective color from the color-guidance image. b) Changing the colors of the color guidance image, results in similar changes in the generated images. c) The exact color which should be generated for the used color name can also be a synthetic example.  }
\label{fig:motivation}
\vspace{-3mm}
\end{figure*}

IP-Adapter~\cite{ye2023ip,rowles2024ipadapterinstruct} has demonstrated remarkable effectiveness in transferring visual attributes from reference images to generated content.
In this paper, we introduce a key insight: the IP-Adapter implicitly establishes latent-space connections between visual features and their semantic counterparts. As illustrated in Figure~\ref{fig:motivation}, this phenomenon, which we call \emph{semantic attribute binding}, creates a direct correspondence between colors in reference images and their linguistic descriptors in prompts. When using a single reference image containing multiple colors (Figure~\ref{fig:motivation}a), the model correctly extracts and applies each color based on the corresponding color name in the prompt (e.g., \quotes{green}, \quotes{yellow} or \quotes{red}). Furthermore, when the colors in the reference image are modified (Figure~\ref{fig:motivation}b), the generated results reflect these changes while maintaining the binding to the appropriate color names. This binding even works with synthetic color patches (Figure~\ref{fig:motivation}c), allowing for precise RGB-level control. These observations reveal that the adapter inherently associates specific color values with their linguistic descriptors, enabling exact color specification without requiring any additional training or fine-tuning.

\noindent\textbf{Color name attribution.}
\emph{Semantic attribute binding} stems from the architectural design of IP-Adapter, particularly its decoupled cross-attention mechanism. As shown by Ye et al.~\cite{ye2023ip}, the IP-Adapter introduces separate cross-attention layers for image features while keeping the original text cross-attention layers intact. Crucially, both cross-attention mechanisms share the same query matrices, while maintaining separate key and value projection matrices for text and image features:
\begin{equation}
\textstyle
    \mathbf{x} = \text{softmax} \left( \frac{\mathbf{Q}\mathbf{K}^\top}{\sqrt{d}} \right) \mathbf{V} + \text{softmax} \left( \frac{\mathbf{Q}\mathbf{K'}^\top}{\sqrt{d}} \right) \mathbf{V'} .
\end{equation}

This sharing of query projection matrices creates an implicit alignment between the text and image feature spaces. During IP-Adapter training, the model learns to map visual color information from the image encoder to the same latent space that the text encoder uses to represent color concepts. Specifically, when a colored image is provided as input, the IP-Adapter learns to bind the visual color properties with their corresponding linguistic color representations that the diffusion model already understands.

To effectively leverage this semantic attribute binding, we need to determine which color names in natural language correspond to specific RGB values. Following foundational work in color linguistics by Berlin and Kay~\cite{berlin1991basic}, we recognize that human languages typically use a limited set of basic color terms to describe the entire color spectrum. In English, there are eleven basic color terms, with eight being chromatic (red, green, blue, yellow, brown, orange, pink, and purple).

To validate our hypothesis that the key projections may be collapsing different color modalities into the same representation, we compute the similarity matrix between color names used in text prompts and RGB color values in reference images. We calculate the projection inner product $\langle \mathbf{K}, \mathbf{K'}\rangle$ to examine similarities between color word tokens from text prompts and RGB feature tokens from the adapter. This similarity measure captures the strength of the semantic binding between textual color descriptors and visual color representations.

\begin{figure}[t]
  \centering
   \includegraphics[width=\linewidth]{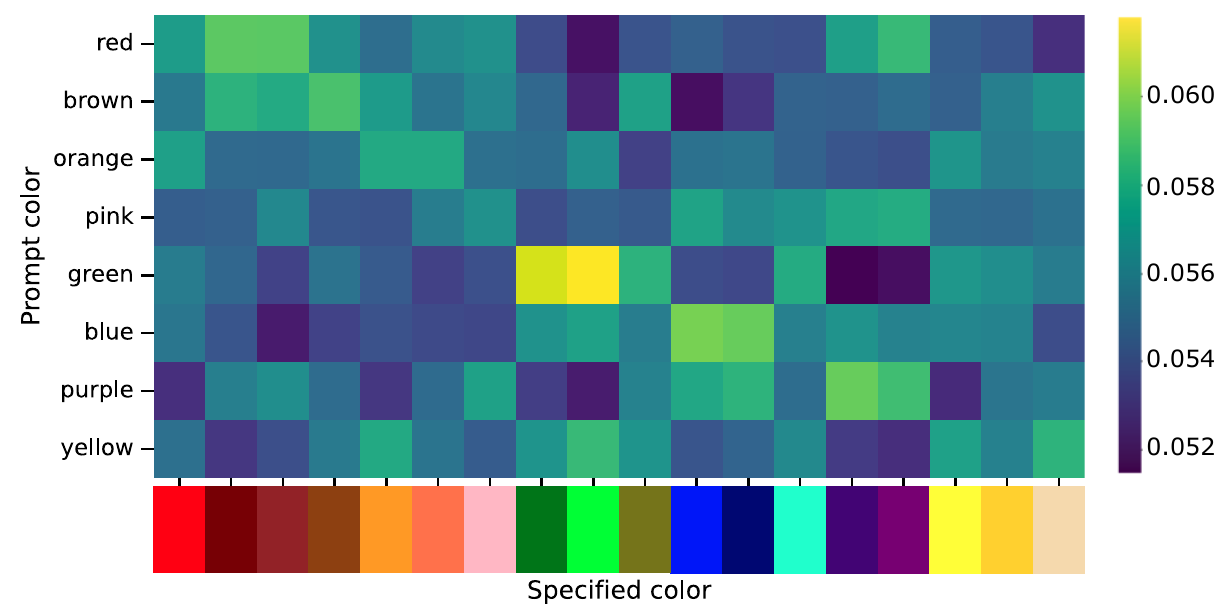}
   \vspace{-7mm}
   \caption{\textbf{Semantic attribute binding}. Phenomenon visualization through a similarity matrix between color names in text prompts and RGB color values. The heatmap shows the normalized dot product similarity between key projections of color word tokens and image features.}
   \vspace{-4.67mm}
   \label{fig:color_name_binding}
\end{figure}

Figure~\ref{fig:color_name_binding} visualizes this similarity matrix for various color names and specified RGB values. The heatmap reveals clear patterns of correspondence, with notably higher similarity values along the diagonal where color names match their expected RGB values (e.g., the ``green'' token shows strongest binding with green RGB values). This confirms our hypothesis that the IP-Adapter implicitly establishes these bindings during its training process, even though it was never explicitly trained for color matching.

Interestingly, we observe that the similarity values are not uniform across all color pairs. Some color terms (like \quotes{green} and \quotes{blue}) show stronger and more distinctive binding patterns, while others exhibit more diffuse relationships. This variation aligns with findings in color linguistics that certain basic color terms have more centralized and consistent representations across languages and visual systems~\cite{kay2006language}.

For automatic color name generation, we use a lookup table that maps arbitrary RGB values to optimal linguistic color descriptors. We construct this using traditional color naming models \cite{van2009learning} to select the color name with maximum similarity. This bidirectional mapping between continuous color space and discrete linguistic categories forms the foundation of our color control approach, allowing us to exploit the implicit color knowledge already encoded in the model's attention mechanisms.\\

\begin{figure}[t]
  \centering
   \includegraphics[width=\linewidth]{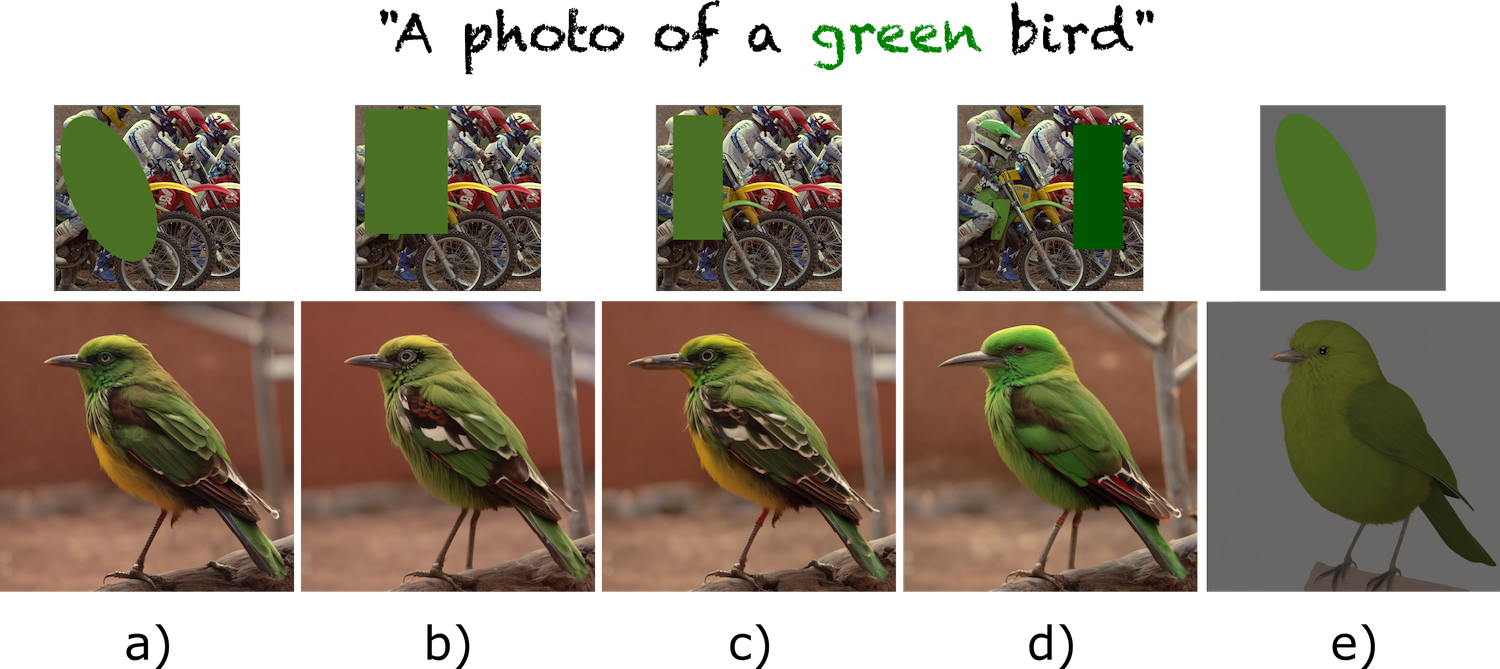}
      \vspace{-5mm}
   \caption{\textbf{Limitations when directly exploiting semantic attribute binding for color control}. (a) Reference image with oval. (b,c) Shape and size variations alter the resulting bird coloration despite using the same green color. (d) With multiple green regions, attribution becomes ambiguous and inconsistent. (e) Using synthetic color references produces flat, unrealistic textures.}
   \vspace{-5mm}
   \label{fig:naive_issues}
\end{figure}

\noindent \textbf{Direct Semantic Attribute Binding Limitations.}
While the binding mechanism offers a promising avenue for color control, naive implementations face several challenges, as illustrated in Figure~\ref{fig:naive_issues}:

\begin{itemize}[leftmargin=*]
    \item \textbf{Shape and Size Sensitivity:} As shown in Figure~\ref{fig:naive_issues}b,c, changing the shape or size of the color reference region alters the resulting object colors. Different image statistics from these variations influence the generation process, leading to inconsistent color attribution.

    \item \textbf{Ambiguous Color Attribution:} When multiple regions contain the desired color (Figure~\ref{fig:naive_issues}d), the binding mechanism struggles to correctly attribute the target color, often selecting one region at random. Furthermore, all colors present in the reference image influence the final appearance of the generated object.

    \item \textbf{Context Limitation:} IP-Adapter generates new images based on the statistics of the reference image, restricting the diversity of backgrounds and contexts (Figure~\ref{fig:naive_issues}e). To generate an object with a specific color in various contexts, each reference image would need to contain that exact color, severely limiting creative freedom.

    \item \textbf{Synthetic Reference Limitation:} Using purely synthetic color references (Figure~\ref{fig:naive_issues}e) results in flat, unrealistic surfaces lacking natural variations, highlights, and textures, producing results that appear artificial and fail to capture the nuanced appearance of real-world objects.

\end{itemize}

To address these challenges, \ourmethod introduces selective attention modulation that preserves generative capabilities while enabling precise color control.

\subsection{Spatial Prior Addition}
In our approach \textit{ColorWave}, we leverage the decoupled cross-attention design established in previous adapters \cite{ye2023ip,wang2024instantstyle} for its minimal interference with pretrained knowledge and compatibility with other methods.
We implement a strategy of minimal yet precise intervention in the generative model to achieve effective color binding. First, we inject the user-specified color information at a single optimal point in the network, where it most effectively influences the desired subject, following successful approaches in prior work \cite{wang2024instantstyle,wang2024instantstyle_plus}. Second, we rely on the model's inherent understanding to determine proper color placement based on its pretrained knowledge.

The challenge is that color information lacks spatial specificity, so we need to ensure it is applied to the correct object. To solve this, we query the model for the target object's location and constrain our intervention specifically to that region, ensuring the specified color binds precisely to the relevant semantic token in the prompt.
As illustrated in Fig.~\ref{fig:overview}, we implement this by masking the adapter's contribution $\mathbf{A}'$ with the attention map of the desired object $\mathbf{A}_{o}$ before incorporating it into the final features $\mathbf{x}$:
\begin{equation} \label{eq:spatial_prior}
    \mathbf{x} = \mathbf{A} + \delta_\lambda(\tilde{\mathbf{A}}_{o}) \mathbf{A}' .
\end{equation}
Here, $\mathbf{A} \in \mathcal{R}^{h \times H \times W}$ represents the original cross-attention output, while $\tilde{\mathbf{A}}_{o}$ is computed by averaging attention maps of the target object across attention heads and $\delta_\lambda$ is a thresholding operator preserving the $\lambda$-th percentile.
This selective approach complements and ensures effective semantic attribute binding described earlier between the specified color and the target object, while minimizing unintended color attribution to other elements in the scene. The evolution of the object mask throughout the generation process is visualized in Section~\ref{sec:maskevolution}.

\section{Experiments}

\subsection{Experimental setup}

\noindent \textbf{Dataset.}
For fair comparison with prior work, we adopt ColorPeel's \cite{butt2025colorpeel} evaluation framework with two color sets: (1) \textbf{Coarse-grained}: four basic colors (red, green, blue, yellow); (2) \textbf{Fine-grained}: eighteen specialized colors (salmon, beige, navy, indigo, etc.), all with predefined RGB values. We use identical prompts (e.g., \quotes{a [color] bowl on the table}) with 20 random seeds each, yielding 200 and 360 images respectively.

Unlike ColorPeel, which requires per-color training, \ourmethod accepts arbitrary RGB values without optimization. The coarse/fine distinction is maintained only for comparison—our method treats all colors uniformly across the RGB spectrum. While evaluation focuses on single-object scenarios, \ourmethod extends to multi-object scenes (see Section~\ref{sec:multiobject}).\\

\noindent \textbf{Evaluation metrics.}
We adopt the evaluation metrics proposed by ColorPeel \cite{butt2025colorpeel}. Specifically, we compute several complementary metrics:

\begin{itemize}
    \item \textbf{Euclidean Distance in CIE Lab color space} ($\Delta_E$ and $\Delta_{E_{Ch}}$ when luminance is removed): These metrics measure perceptual uniformity between generated and target colors; lower values indicate better color matching.

    \item \textbf{Mean Angular Error (MAE) in sRGB}: quantifies color deviation in terms of chromaticity, helping to understand differences in hue and saturation independent of intensity.

    \item \textbf{Mean Angular Error (MAE) in Hue}: This analyzes the difference between target and generated colors irrespective of brightness and saturation.
\end{itemize}
For metric and implementation details refer to Section~\ref{sec:experimentaldetails}.\\

\begin{figure}[t]
  \centering
   \includegraphics[width=1\linewidth]{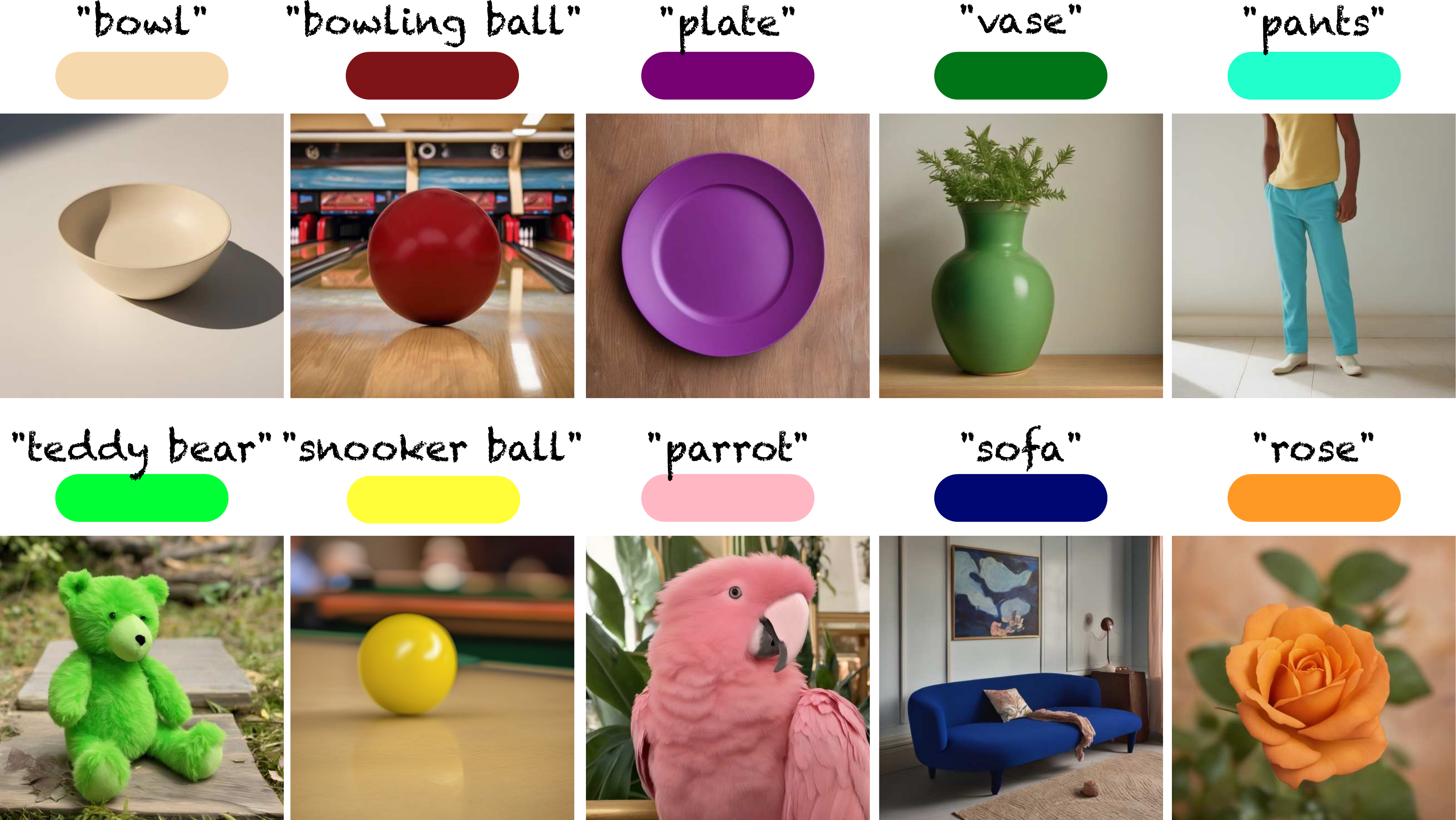}
   \vspace{-5mm}
   \caption{\textbf{Arbitrary color attribution across diverse subjects}. \ourmethod precisely applies user-specified colors to various target objects while maintaining natural lighting, material properties, and contextual integration.}
   \label{fig:color_galore}
   \vspace{-4mm}
\end{figure}

\noindent \textbf{Comparison methods.}
\ourmethod represents the first training-free approach for precise color control in diffusion models, establishing a distinct methodological category from existing approaches that require color-specific training. Our primary comparison is with ColorPeel \cite{butt2025colorpeel}, the current state-of-the-art in color-specific prompt learning.
For completeness, we compare against all baselines evaluated in ColorPeel, that includes \textit{training-free} T2I generation baselines: (1) vanilla Stable Diffusion with color name text prompts; (2) Rich-Text \cite{ge2023richtext} and Rich-Text++~\cite{ge2025expressive}, which enhances adherence to complex text descriptions;
And also \textit{training-based} personalization methods besides ColorPeel~\cite{butt2025colorpeel}: (3) Textual Inversion \cite{textual_inversion}, learning new pseudo-words in embedding space; (4) DreamBooth \cite{ruiz2022dreambooth}, fine-tuning the entire diffusion model; and (5) Custom Diffusion \cite{kumari2022customdiffusion}, optimizing projection matrices in cross-attention layers.
The training-free nature of \ourmethod creates a fundamental asymmetry in this comparison – while ColorPeel and other training-based approaches require separate optimization processes for each individual color or complex training regimes to handle multiple colors simultaneously, \ourmethod inherently accepts any arbitrary RGB triplet without modification. This represents not just an incremental improvement but a paradigm shift in how to achieve precise color control in generative models, transcending the limitations of discrete color vocabularies and model-specific optimizations.

\subsection{Qualitative Comparisons}
To demonstrate the versatility and effectiveness of \textit{ColorWave}, we present qualitative results across several key capabilities: precise color attribution across diverse objects, fine-grained color control, and generalization to complex color patterns and textures. These experiments highlight the flexibility of our training-free approach in scenarios that would require extensive additional training in previous methods. More results can be found in Section~\ref{sec:extraresults}.

\noindent \textbf{Arbitrary color attribution.}
We evaluate \ourmethod's ability to precisely apply user-specified colors to a wide range of subjects including inanimate objects, animals, plants, and human clothing. Figure \ref{fig:color_galore} demonstrates that our method successfully generates images where target objects accurately reflect the desired reference colors. Notably, \ourmethod maintains high visual quality across diverse scenarios
while achieving precise color matching.

\noindent \textbf{Fine color navigation.}
Beyond coarse color attribution, we examine whether our method can detect and reproduce subtle color variations. Figure \ref{fig:color_interpolation} showcases results from smooth interpolation between similar color tones.
This precision demonstrates the sensitivity of semantic attribute binding and shows that the proposed method can navigate the continuous color space with remarkable accuracy.

\begin{figure}[t]
  \centering
   \includegraphics[width=1\linewidth]{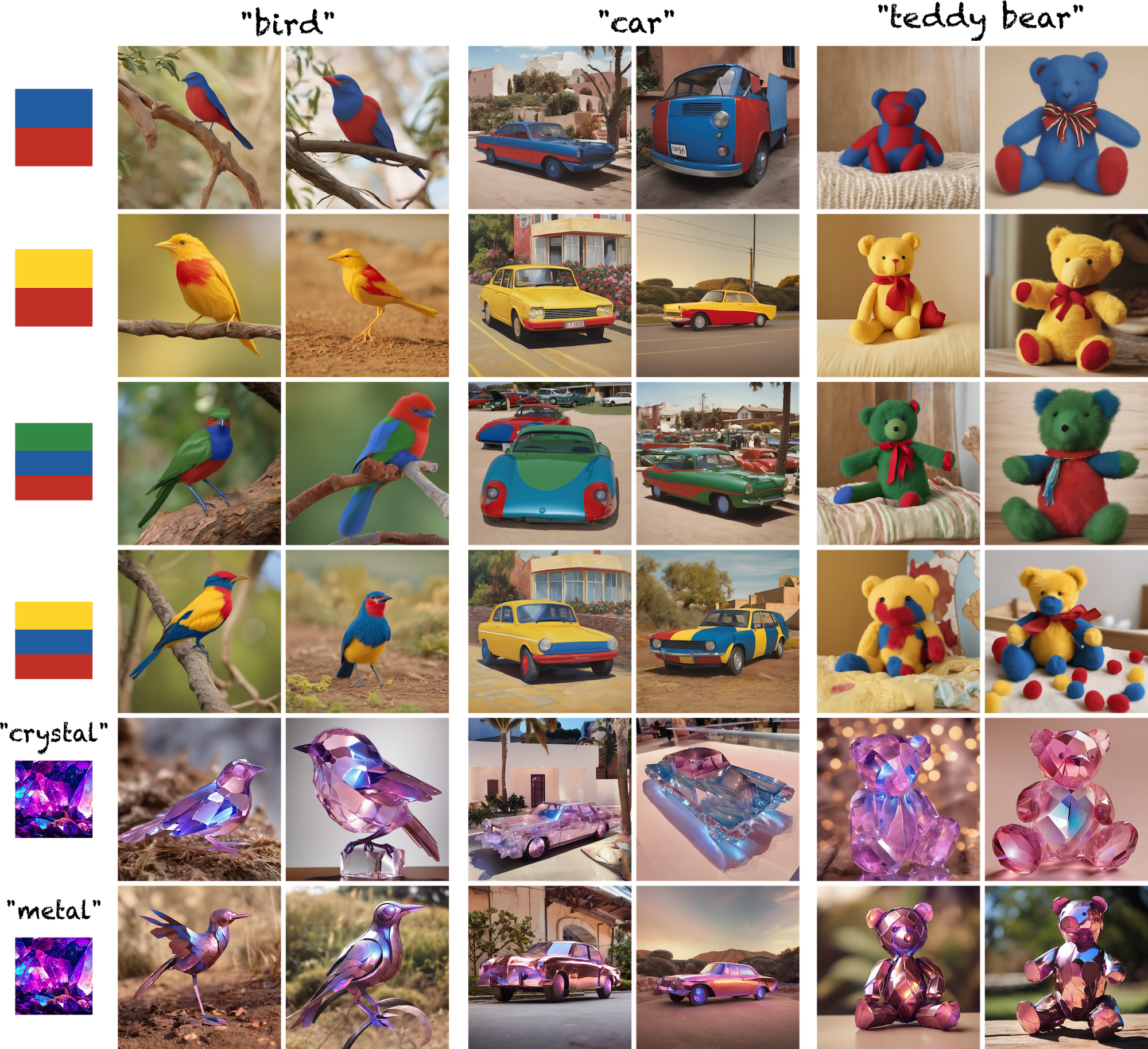}
   \vspace{-6mm}
  \caption{\textbf{Generalizability to complex color patterns and textures}. These examples illustrate how semantic attribute binding extends beyond simple color matching to more sophisticated visual attribute control.}
   \vspace{-5mm}
   \label{fig:generalizability}
\end{figure}

\noindent  \textbf{\ourmethod generalizability.}
We further investigate the extensibility of our approach beyond single-color attribution. Figure \ref{fig:generalizability} demonstrates two advanced applications: (1) applying complete color palettes to target objects, and (2) transferring complex textures while maintaining precise color control. In both cases, the method preserves the structural integrity and material properties of the target objects while successfully integrating the specified color patterns.

\subsection{Quantitative Analysis}

We compare \ourmethod with training-based and training-free baselines. Such quantitative results are demonstrated in Table~\ref{tab:Quantitative_comparison}.
For each method, we generated images, and extract the mask of the object, following the same pipeline as in ColorPeel~\cite{butt2025colorpeel}. Percentages in MAE metrics denote the percentage of pixels inside the mask used for the computation (selecting those closest to the ground truth).
This table clearly shows the superiority of our method. Particularly, \ourmethod achieved notably lower $\Delta E$ error in CIE Lab color space as compared to the existing training-free methods and it is only worse than the training-based state-of-the-art ColorPeel, which indicates that \ourmethod generates perceptually better colors than its direct competitors.
In addition, \ourmethod also achieved comparatively much lower mean angular error in both sRGB and Hue when compared to the other training-free methods, which signifies a higher degree of color accuracy in terms of chromaticity and hue in our generated images.
While ColorPeel achieves superior numerical performance on some metrics due to its color-specific training, our training-free approach offers significant practical advantages: immediate RGB specification without optimization, broader architectural compatibility, and elimination of per-color training requirements.\\

\begin{table*}[t]
    \centering
    \small
    \caption{Quantitative comparison with baselines over various evaluation metrics. All metrics the smaller the better ($\downarrow$). The best results of both \textit{training-based} and \textit{training-free} technique streams are highlighted in bold. Training time is provided in the last column.
    }
    \vspace{-2mm}
    \label{tab:Quantitative_comparison}
    \resizebox{0.84321\textwidth}{!}{
    \begin{tabular}{ c  c | c | c | ccc | ccc | c }
    \toprule
    \multicolumn{2}{c|}{\multirow{2}{*}{{Method}}}  &  \multirow{2}{*}{{$\Delta E$}} 	&   \multirow{2}{*}{{$\Delta E_{ch}$}} 	 & \multicolumn{3}{c|}{MAE (sRGB)}	&  \multicolumn{3}{c|}{MAE (Hue)}  & Time \\

    & & & & 10\% & 50\% & 100\% & 10\% & 50\% & 100\% & (min) \\
    \midrule
    \multirow{7}{*}{{\rotatebox{90}{\makecell{Training \\ Based}}}}& TI~\cite{textual_inversion} & 48.98 & 44.29 & 15.22 & 19.51 & 23.90 & 52.66 & 69.35 & 90.88 & 118  \\
    & DB~\cite{ruiz2022dreambooth} & 50.71 & 46.29 & 14.75 & 19.30 & 23.70 & 47.12 & 67.13 & 88.72 & 56  \\
    & CD~\cite{kumari2022customdiffusion} & 48.47 & 42.23 & 13.43 & 17.93 & 22.43 & 31.63 & 55.07 & 78.43 & 24  \\
    & ColorPeel (3D)~\cite{butt2025colorpeel} &  {21.39} &	 {16.51}	& \textbf{4.36}	& \textbf{7.76} &	\textbf{12.08} &	\textbf{2.63} & \textbf{6.47} & \textbf{21.35} & 19 \\
    & ColorPeel (2D)~\cite{butt2025colorpeel} & \textbf{20.45} &	\textbf{15.29}	& {4.83}	&  {7.88} &	 {12.13} &	 {3.18} &  {7.43} &  {21.46} & 19\\
    & ColorPeel (3D)~\cite{butt2025colorpeel} \scriptsize{SD-XL} & 29.01 & 23.65 & 5.46 & 8.19 & 12.59 & 9.98 & 26.28 & 51.71 & 41 \\
    & ColorPeel (2D)~\cite{butt2025colorpeel} \scriptsize{SD-XL} & 28.56 & 23.25 & 5.31 & 8.06 & 12.45 & 9.28 & 25.24 & 50.53 & 41 \\
    \midrule
    \multirow{6}{*}{{\rotatebox{90}{\makecell{Training \\ Free}}}} & SD~\cite{Rombach_2022_CVPR_stablediffusion} & 47.45 & 41.55 & 12.89 & 20.04 & 26.93 & 30.17 & 54.14 & 86.38 & {0} \\
    & Rich-Text~\cite{ge2023richtext} & 36.62 & 32.48 & 9.91 &  13.29 & 18.53 & 50.55 & 72.77 & 93.51 & {0} \\
    & Rich-Text++~\cite{ge2025expressive} \scriptsize{SD-XL} & 60.31 & 57.04 & 21.42 & 24.35 & 27.13 & 81.33 & 115.08 & 139.72 & {0} \\
    & \ourmethod (Ours) \scriptsize{SD-XL} & 22.81 & 19.64 & 5.89 & 7.83 & \textbf{11.00} & 14.94 & 23.12 & 39.68 & {0} \\
    & \ourmethod (Ours) \scriptsize{SD3.5-L} & \textbf{16.04} & \textbf{12.25} & \textbf{5.01} & \textbf{7.78} & 11.05 & \textbf{3.05} & \textbf{8.44} & \textbf{27.80} & {0} \\
    & \ourmethod (Ours) \scriptsize{FLUX.1-dev} & 23.04 & 18.91 & 5.64 & 8.64 & 12.10 & 10.57 & 27.96 & 49.20 & {0} \\
    \bottomrule
    \end{tabular}
    \vspace{-6mm}
    }
\end{table*}

\begin{figure}[t]
  \centering
   \includegraphics[width=0.99\linewidth]{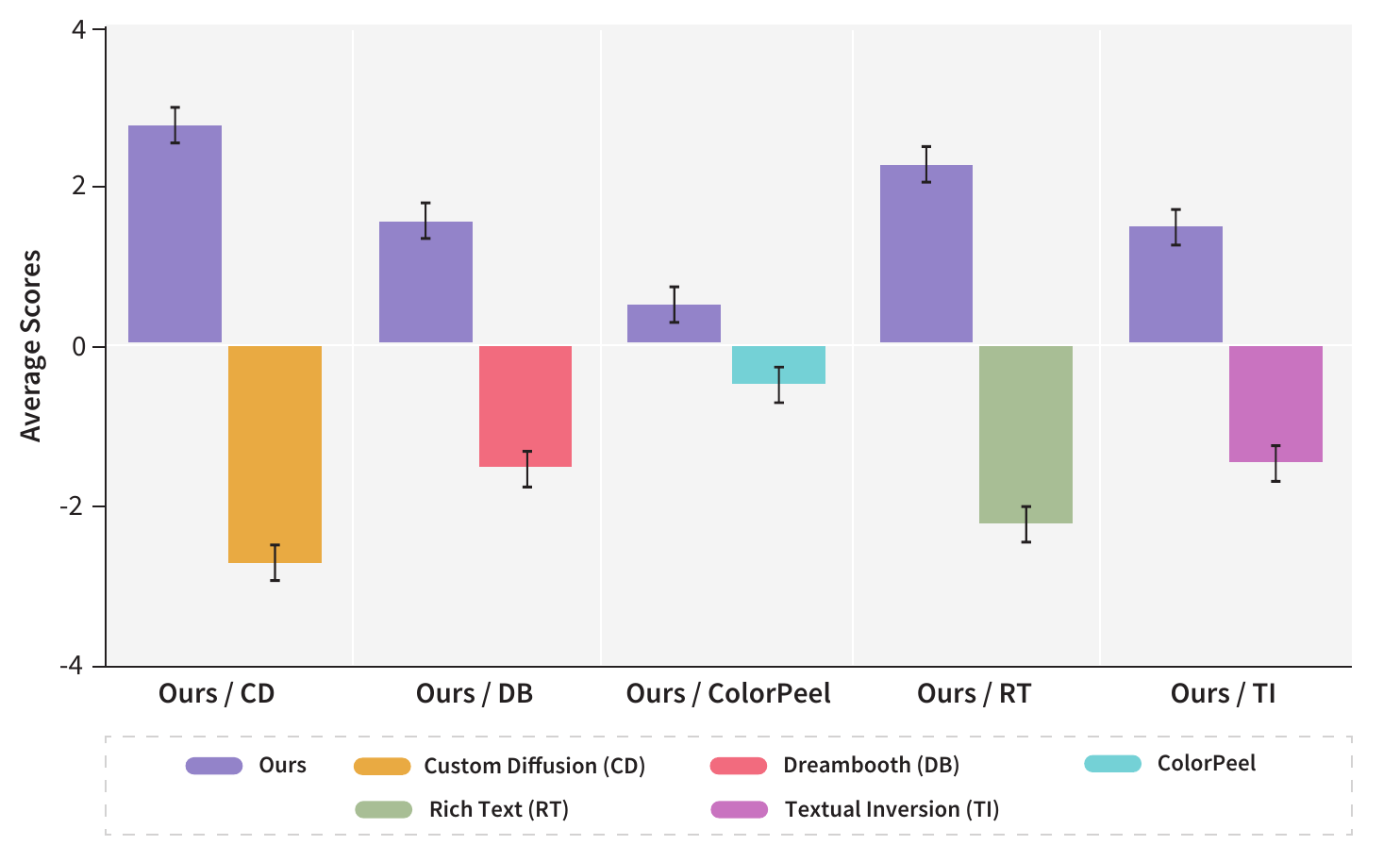}
   \vspace{-1mm}
   \caption{
   \textbf{User study}. Z-scores are higher for better human preference. Our method outperforms baselines and the state-of-the-art color control generation method ColorPeel~\cite{butt2025colorpeel}.
   }
   \vspace{-4mm}
   \label{fig:user_study}
\end{figure}

\noindent \textbf{User study.}
We conducted a user study with 25 participants to perceptually evaluate our results, comparing \ourmethod against ColorPeel~\cite{butt2025colorpeel}, TI~\cite{textual_inversion}, Rich-Text~\cite{ge2023richtext}, DB~\cite{ruiz2022dreambooth}, and CD~\cite{kumari2022customdiffusion}. We followed the same experimental paradigm as the one defined in ColorPeel.
The experiment took place in a controlled lab environment to ensure reliability. All participants were tested for correct color vision using the Ishihara test. The study followed a two-alternative forced choice (2AFC) method. Observers viewed three images on a monitor set to RGB: the central image represented the target color, while the left and right images displayed results generated by our method and one of the competing methods, with their positions randomized. We tested the same 10 different prompts and four colors (red, green, blue, and yellow) defined in ColorPeel.
To analyze the results, we compared \ourmethod against each competing method using the Thurstone Case V Law of Comparative Judgment model \cite{thurstone2017}. This approach provided us with z-scores and a 95$\%$ confidence interval, calculated using the method proposed in \cite{montag2006}. The results, shown in Fig.~\ref{fig:user_study}, indicate that \ourmethod is statistically significantly better than all five competing algorithms, including ColorPeel, which learns a specific token for each color. These findings highlight \ourmethod's effectiveness in generating more realistic and accurate colors given an RGB triplet.

\subsection{Ablations}

\begin{figure}[t]
  \centering
   \includegraphics[width=\linewidth]{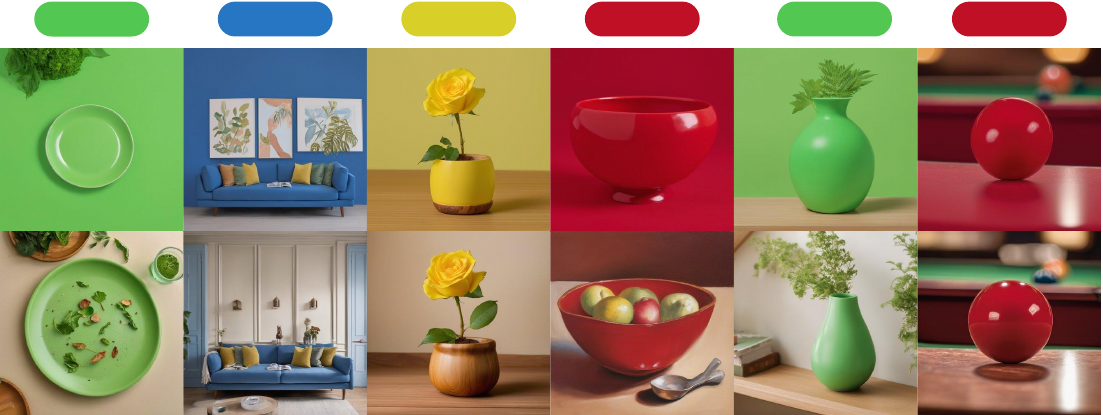}

  \vspace{-1mm}
   \caption{\textbf{Comparison between direct IP-Adapter usage and \textit{ColorWave}}. (top) Naive IP-Adapter exhibits flat appearance and significant color leakage. (bottom) \ourmethod demonstrates realistic lighting, shadows, and material properties while maintaining precise color control without leakage.}
   \vspace{-3mm}
   \label{fig:ablation_comparison_ipa}
\end{figure}

\noindent \textbf{Comparison to IP-Adapter.}
Direct IP-Adapter application for color control using solid color patches or colored objects suffers from significant limitations that \ourmethod addresses.
Figure~\ref{fig:ablation_comparison_ipa} and Table~\ref{tab:ablation_comparison_ipa} demonstrate naive IP-Adapter usage exhibits substantial color leakage (36.03\% of non-target pixels affected vs. 8.80\% for \ourmethod), where colors bleed into unintended regions beyond the target object, and produces flat, unrealistic appearances lacking natural lighting variations, shadows, and material properties. The leakage metric definition can be found in Section~\ref{sec:colorleakmetric}. Additionally, direct IP-Adapter generates lower quality images as measured by no-reference metrics NIQE~\cite{6353522} (6.17 vs. 4.78) and BRISQUE~\cite{6272356} (29.76 vs. 21.49).
However, to address the photorealism limitations inherent in direct IP-Adapter application, our method introduces spatial prior addition (Equation~\ref{eq:spatial_prior}), which constrains color application through attention map masking, preventing unwanted leakage and preserving natural lighting variations, shadows, and material properties. Despite these improvements in spatial control and visual realism, \ourmethod maintains similar color accuracy to the direct approach, demonstrating that our method successfully addresses IP-Adapter's fundamental limitations for precise color control tasks while enhancing overall generation quality.
\begin{table}[t]
\centering
\small
\caption{Comparison between direct IP-Adapter usage and \ourmethod for color control tasks. All metrics the smaller the better ($\downarrow$).}
\vspace{-2mm}
\label{tab:ablation_comparison_ipa}
\begin{tabular}{@{}lccc@{}}
\toprule
\multicolumn{1}{c}{Method} & \multicolumn{1}{c}{\begin{tabular}[c]{@{}c@{}}Color\\ leakage\end{tabular}} & \multicolumn{1}{c}{NIQE} & \multicolumn{1}{c}{BRISQUE} \\ \midrule
IP-Adapter                 & 36.03                                                                       & 6.17                     & 29.76                       \\
\ourmethod                  & 8.80                                                                        & 4.78                     & 21.49                       \\ \bottomrule
\end{tabular}
\end{table}

\begin{figure}[t]
  \centering
   \includegraphics[width=\linewidth]{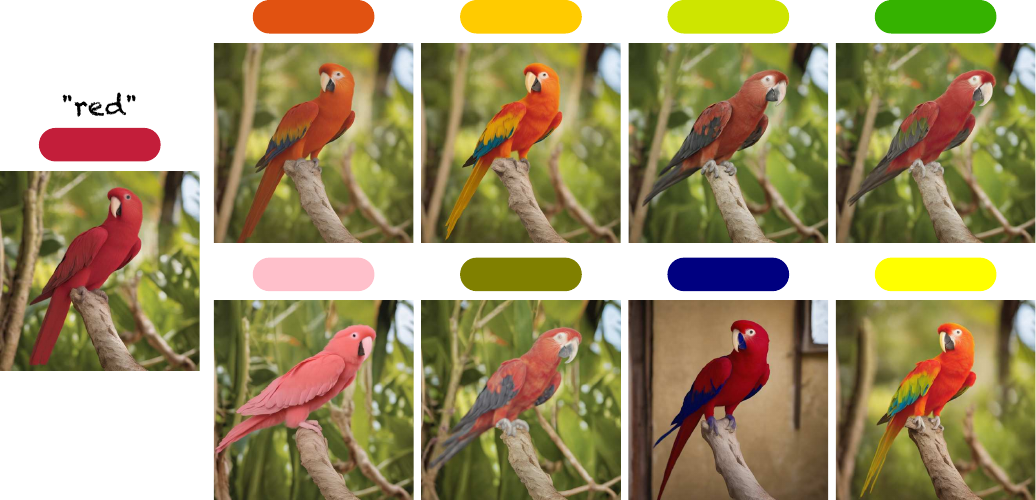}

   \caption{\textbf{Boundaries of semantic attribute binding when prompt and target colors diverge}. All images were generated using the prompt \textit{``A photo of a \textcolor{red}{red} parrot''} while varying the input color. This demonstrates that semantic attribute binding operates within color neighborhoods, with effectiveness decreasing as perceptual distance increases between prompt color and target color.}
   \vspace{-4mm}
   \label{fig:ablation_cn_attrib}
\end{figure}

\noindent \textbf{Color name attribution.}
We investigate semantic attribute binding constraints by creating discrepancies between prompt colors and specified input colors. Figure \ref{fig:ablation_cn_attrib} shows results using consistent prompt \textit{``A photo of a red parrot''} while varying input colors.
Our findings reveal a semantic proximity effect, understood by Fig.~\ref{fig:color_name_binding}; the success of color attribution depends on the perceptual and semantic distance between the prompt color and the target color. As shown in Figure \ref{fig:ablation_cn_attrib} top, when the target color (orange) shares perceptual similarity with the prompt color (red), our method successfully binds the new color to the parrot. However, when the target color (green) represents an opponent color on the color wheel, the binding fails, and the model defaults to generating a parrot closer to the prompted red color.
The lower row demonstrates this effect across a broader range of colors.
Pink succeeds due to shared properties with red (both are warm colors with similar hue angles), while semantically distant colors (olive, navy, yellow) fail to override red descriptors.

\section{Conclusion}
We introduced \textit{ColorWave}, the first training-free method for precise color control in text-to-image diffusion models. By exploiting latent links between visual attributes and semantic cues in IP-Adapter, our approach enables direct RGB-level manipulation without fine-tuning. It rewires cross-attention to achieve accurate color attribution while preserving generation quality. Extensive evaluations show \ourmethod outperforms training-free methods in color accuracy and rivals fine-tuned approaches like ColorPeel—offering faster, more efficient alternatives to prompt-based color control. This reveals new, untapped capabilities in existing diffusion models.

\footnotesize{
\noindent \textbf{Acknowledgments.} This work was supported by Grants PID2021-128178OB-I00, PID2022-143257NB-I00, and PID2024-162555OB-I00 funded by MCIN/AEI/10.13039/ 501100011033 and FEDER, by the Generalitat de Catalunya CERCA Program, by the grant Catedra ENIA UAB-Cruïlla (TSI-100929-2023-2) from the Ministry of Economic Affairs and Digital Transition of Spain. JVC also acknowledges the 2025 Leonardo Grant for Scientific Research and Cultural Creation from the BBVA Foundation. The BBVA Foundation accepts no responsibility for the opinions, statements and contents included in the project and/or the results thereof, which are entirely the responsibility of the authors. Kai Wang acknowledges the funding from Guangdong and Hong Kong Universities 1+1+1 Joint Research Collaboration Scheme and the start-up grant B01040000108 from CityU-DG. The authors thankfully acknowledges RES resources provided by BSC in MareNostrum5 to IM-2025-3-0025, IM-2025-3-0027.
}

{
    \small
    \bibliographystyle{ieeenat_fullname}
    \bibliography{main}
}

\clearpage
\appendix

\section{Color binding regions as a function of steps} \label{sec:maskevolution}

\begin{figure}[ht]
  \centering
   \includegraphics[width=0.85\linewidth]{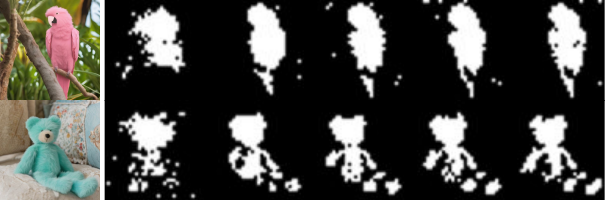}
   \vspace{-2mm}
   \caption{Color binding regions based on $\delta_\lambda(\tilde{\mathbf{A}}_{o})$. Depicted every 10-th denoising step.}
   \vspace{-2mm}
   \label{fig:ip_mask}
\end{figure}

Figure \ref{fig:ip_mask} displays color binding regions based on $\delta_\lambda(\tilde{\mathbf{A}}_{o})$ every 10-th denoising step,

\section{Multi-object} \label{sec:multiobject}

\begin{figure}[t]
  \centering
   \includegraphics[width=\linewidth]{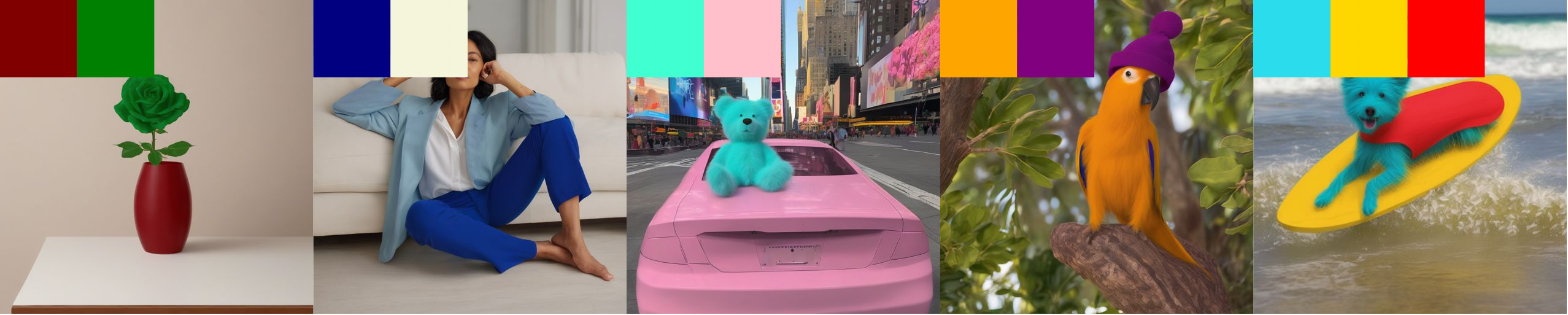}
   \caption{
   Multi-object color control examples. (first) ``\textit{A green rose growing from a red vase on a table}'', (second) ``\textit{A woman wearing blue pants sitting on a white sofa}'', (third) ``\textit{A blue teddy bear in Times Square on top of a pink car}'', (fourth) ``\textit{An orange parrot with a purple hat perched on a tree}'', (fifth) ``\textit{A photo of a blue dog on a yellow surfboard surfing a wave wearing a orange lifevest}''.
   }
   \label{fig:multi_obj}
\end{figure}

ColorWave successfully applies different colors to multiple objects within the same scene by leveraging separate attention maps for each target object token, as demonstrated in Figure~\ref{fig:multi_obj}. As seen, the method achieves a robust color control. Potential failure cases may arise when the object has faulty or misrepresented attention maps, therefore rendering the color binding difficult or unfaithful.

\section{Experimental details} \label{sec:experimentaldetails}

\noindent \textbf{Evaluation metric details}
For each generated image, we use the Segment-Anything model \cite{kirillov2023segment} to automatically extract masks for the target objects, allowing precise measurement of color attributes only within relevant regions. We report these metrics under multiple conditions: considering all pixels within the object mask (100\%), and also considering the top 10\% and 50\% of pixels closest to the target color, which helps account for natural variations in object appearance such as highlights and shadows.\\

\noindent \textbf{Implementation details}
We implement \ourmethod using a pretrained Stable Diffusion XL (SDXL) \cite{podell2023sdxl} model as our base architecture. For the image conditioning component, we utilize the IP-Adapter framework \cite{ye2023ip} with an encoder based on  OpenCLIP-ViT-H-14 \cite{Ilharco_Open_Clip_2021}.
We only inject the color embedding into \textit{the first decoder layer} to compute cross-attention maps, which shows better \textit{stylization} performance as previous works proved~\cite{wang2024instantstyle,wang2024instantstyle_plus,agarwal2023image_matte}.

For the adapter masking, we keep the top 20\% largest values on each object map.
During inference, we process the user-specified RGB values  that creates temporary color reference images. These references are encoded by the IP-Adapter and strategically injected into the model's cross-attention layers. \\

\section{Color Leakage Metric} \label{sec:colorleakmetric}

To quantify unintended color attribution in generated images, we introduce a color leakage metric that measures the percentage of pixels in non-target regions that exhibit colors similar to the specified target color. This metric is particularly important for evaluating whether color control methods successfully confine the desired color to the intended object without affecting other parts of the image.

The color leakage metric operates by first extracting the reference RGB color from the image filename, then converting both the reference color and the generated image to HSV color space. For each pixel in the image, we calculate the circular hue difference between the pixel's hue and the reference hue, accounting for the wraparound nature of the hue channel at $180^{\circ}$. Pixels with hue differences within a threshold of 10 degrees are considered to match the target color.

\section{Mechanism of Semantic Attribute Binding}

\begin{figure*}[ht]
  \centering
   \includegraphics[width=0.85\linewidth]{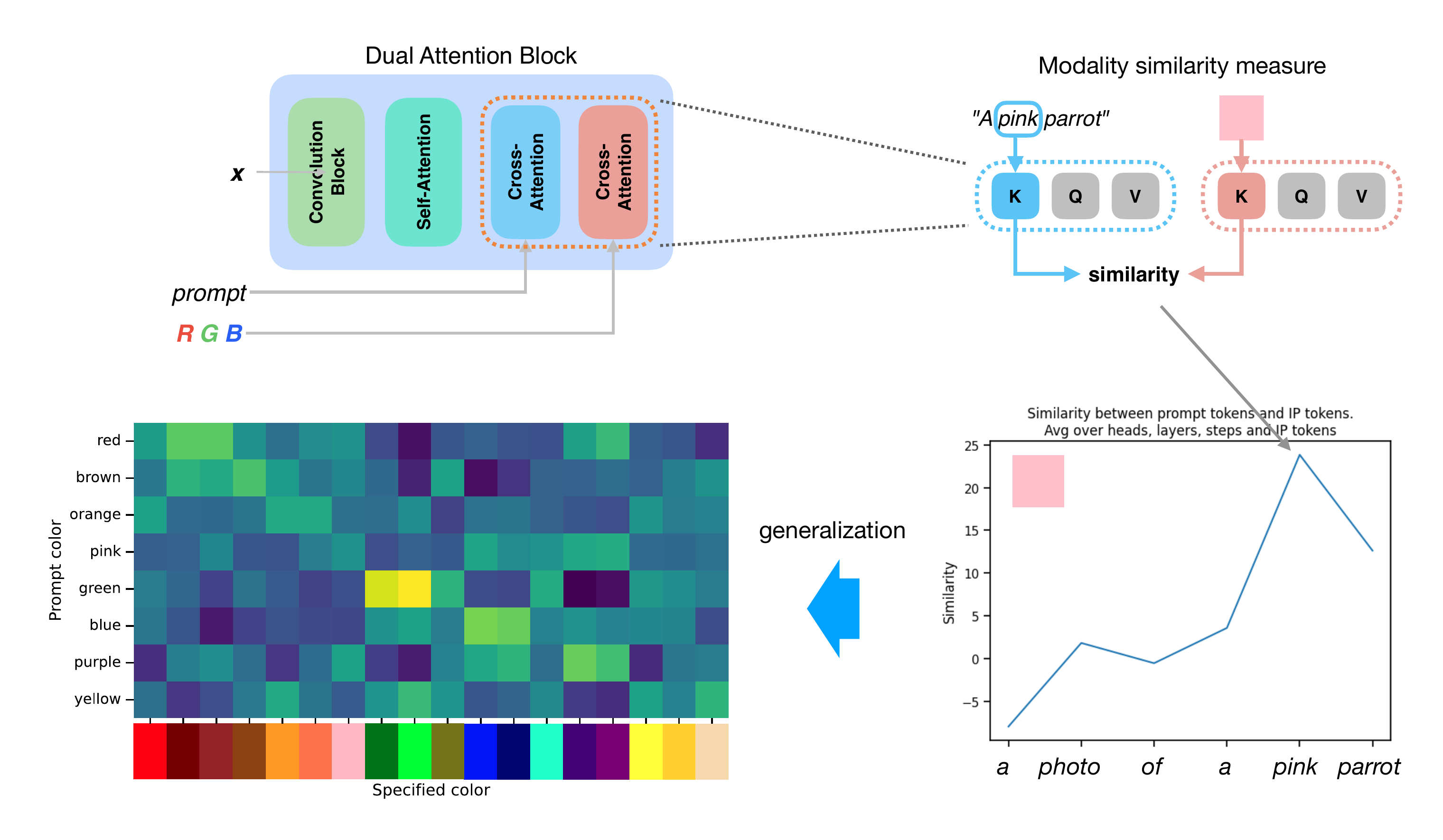}
   \caption{The dual cross-attention architecture enables implicit alignment between text and image features through shared query projections. The similarity heatmap (bottom left) shows correspondence patterns between prompt colors and target colors, with diagonal peaks indicating successful color-semantic binding. The token-level similarity plot (bottom right) demonstrates how individual prompt tokens relate to image features, with color words showing peak correspondence to matching visual color information.}
   \label{fig:mechanism}
\end{figure*}

As depicted in Figure~\ref{fig:mechanism}, the semantic attribute binding phenomenon arises from IP-Adapter's decoupled cross-attention design, where shared query projection matrices create implicit alignment between visual and textual feature spaces. We quantify this binding by computing the inner product similarity $\langle \mathbf{K}_\text{text}, \mathbf{K}_\text{image} \rangle$ between text prompt tokens and image feature tokens.

The heatmap reveals systematic correspondence patterns across color combinations, with diagonal peaks indicating that the model implicitly learns to associate linguistic color terms with their visual counterparts. The token-level analysis demonstrates that color words (e.g., "pink") exhibit peak similarity with corresponding visual color features, while non-color tokens maintain lower similarities. This mechanism generalizes across the continuous color space, with perceptually similar colors showing elevated cross-similarities that reflect the model's understanding of color relationships.

\section{Extra results of \ourmethod} \label{sec:extraresults}

\begin{figure*}[ht]
  \centering
   \includegraphics[width=0.85\linewidth]{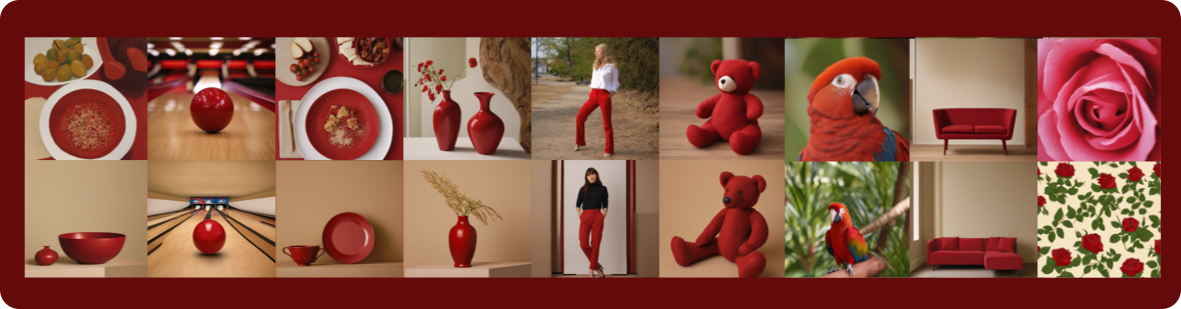}
   \caption{"A photo of a red -object-" - Reference color is depicted in the external frame of the grid - Objects same as in ColorPeel method (see qualitative evaluation)}
   \label{fig:sup_red}
\end{figure*}

\begin{figure*}[t]
  \centering
   \includegraphics[width=0.85\linewidth]{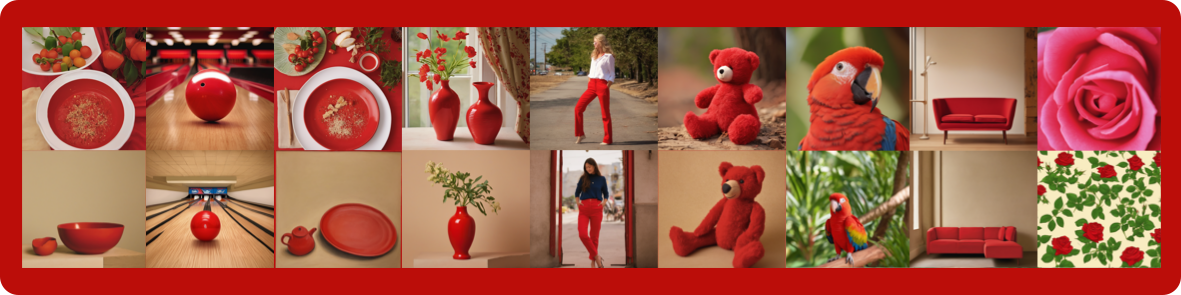}
   \caption{"A photo of a red -object-" - Reference color is depicted in the external frame of the grid - Objects same as in ColorPeel method (see qualitative evaluation)}
   \label{fig:sup_red2}
\end{figure*}

\begin{figure*}[t]
  \centering
   \includegraphics[width=0.85\linewidth]{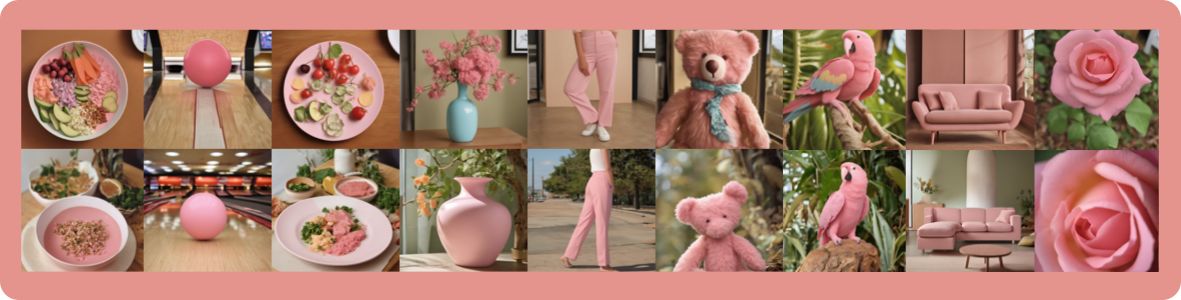}
   \caption{"A photo of a pink -object-" - Reference color is depicted in the external frame of the grid - Objects same as in ColorPeel method (see qualitative evaluation)}
   \label{fig:sup_red3}
\end{figure*}

\begin{figure*}[t]
  \centering
   \includegraphics[width=0.85\linewidth]{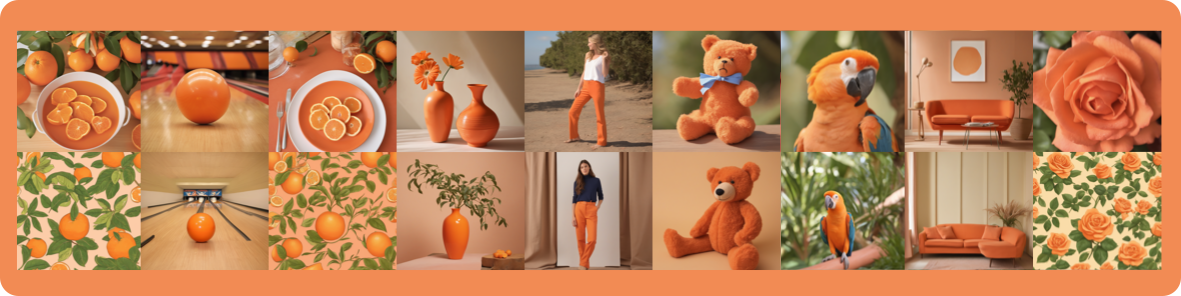}
   \caption{"A photo of a orange -object-" - Reference color is depicted in the external frame of the grid - Objects same as in ColorPeel method (see qualitative evaluation)}
   \label{fig:sup_red4}
\end{figure*}

\begin{figure*}[t]
  \centering
   \includegraphics[width=0.85\linewidth]{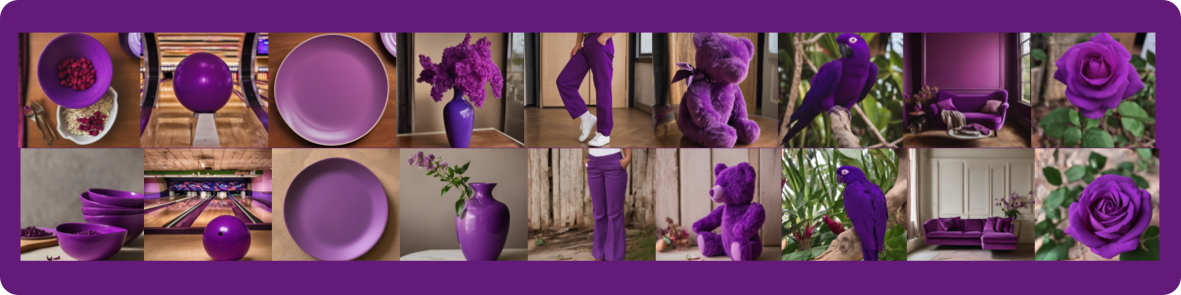}
   \caption{"A photo of a purple -object-" - Reference color is depicted in the external frame of the grid - Objects same as in ColorPeel method (see qualitative evaluation)}
   \label{fig:sup_red5}
\end{figure*}

\begin{figure*}[t]
  \centering
   \includegraphics[width=0.85\linewidth]{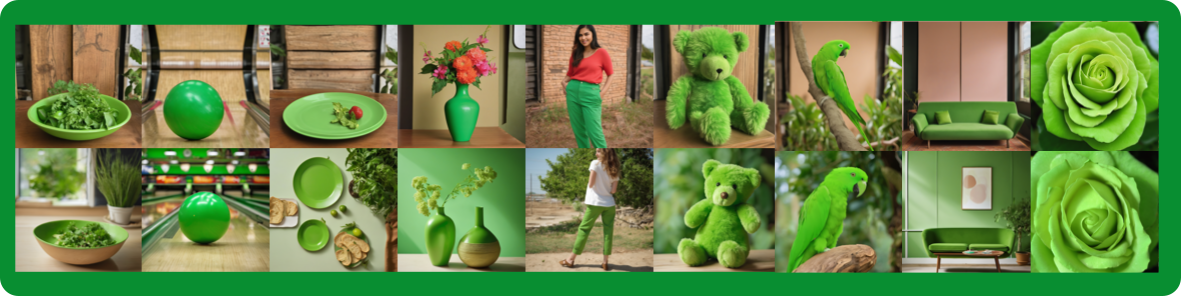}
   \caption{"A photo of a green -object-" - Reference color is depicted in the external frame of the grid - Objects same as in ColorPeel method (see qualitative evaluation)}
   \label{fig:sup_red6}
\end{figure*}

\begin{figure*}[t]
  \centering
   \includegraphics[width=0.85\linewidth]{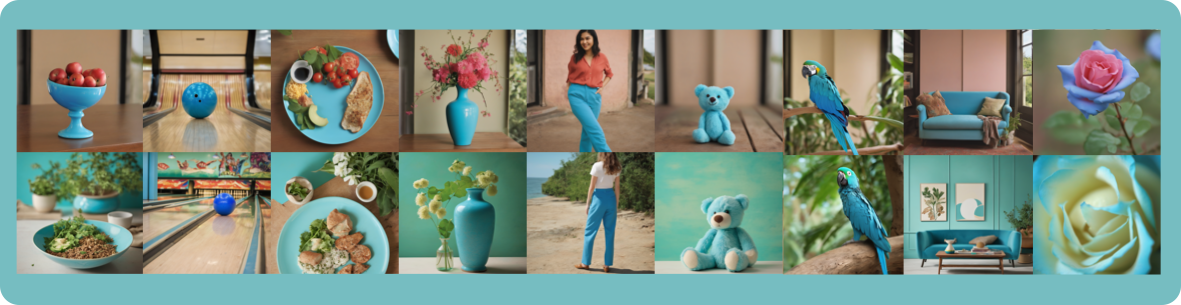}
   \caption{"A photo of a blue -object-" - Reference color is depicted in the external frame of the grid - Objects same as in ColorPeel method (see qualitative evaluation)}
   \label{fig:sup_red7}
\end{figure*}

\begin{figure*}[t]
  \centering
   \includegraphics[width=0.85\linewidth]{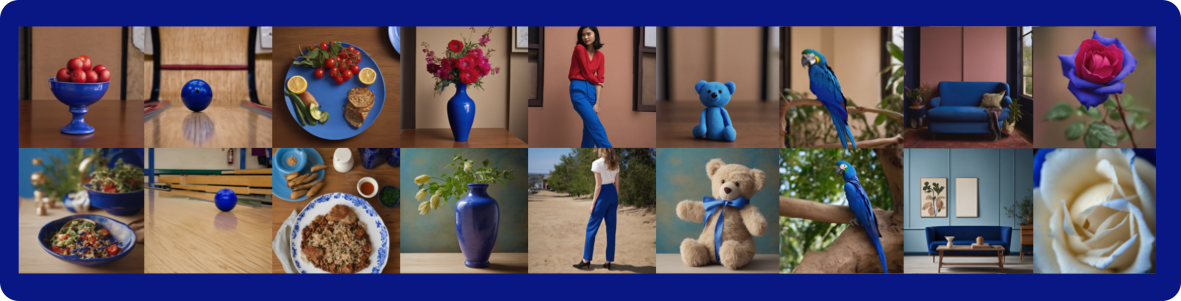}
   \caption{"A photo of a blue -object-" - Reference color is depicted in the external frame of the grid - Objects same as in ColorPeel method (see qualitative evaluation)}
   \label{fig:sup_red8}
\end{figure*}

Extra results are showcased in Figures \ref{fig:sup_red} to \ref{fig:sup_red8}.

\end{document}